\documentclass{iauc}
\usepackage{graphicx}

\title[Terrestrial Exoplanet Light Curves] 
{Terrestrial Exoplanet Light Curves}

\author[Gaidos, Moskovitz \& Williams]   
{Eric Gaidos$^1$, Nicholas Moskovitz$^2$ \and Darren M. Williams$^3$}

\affiliation{$^1$School of Ocean/Earth Sciences \& Technology, University of Hawaii,
Honolulu, HI 96822 USA\break 
email: gaidos@hawaii.edu\\[\affilskip]
$^2$Institute for Astronomy, University of Hawaii, Honolulu, HI 96822 USA\break
email: nmosko@ifa.hawaii.edu\\[\affilskip]
$^3$School of Science, Penn State Erie, The Behrend College, Erie, PA 16563 USA\break 
email: dmw145@psu.edu}

\pubyear{2006}
\volume{200} 
\pagerange{xxx--xxx}
\date{?? and in revised form ??}
\jname{Direct Imaging of Exoplanets: Science \& Techniques}
\editors{C. Aime and F. Vakili, eds.}
\begin{document}

\maketitle

\begin{abstract}
The phase or orbital light curves of extrasolar terrestrial planets in
reflected or emitted light will contain information about their
atmospheres and surfaces complementary to data obtained by other
techniques such as spectrosopy.  We show calculated light curves at
optical and thermal infrared wavelengths for a variety of Earth-like
and Earth-unlike planets. We also show that large satellites of
Earth-sized planets are detectable, but may cause aliasing effects if
the lightcurve is insufficiently sampled.  

\end{abstract}

\firstsection 
\section{Introduction}

The variation in the reflected or emitted light from a planet as it
orbits its parent star contains unique information about that body.
In 1610, Galileo Galilei used his telescope to discover that the
changing brightness of Venus was partly a consequence of its phases
(and also overthrowing the Ptolmeic theory of the Cosmos).  His
anagram to the Tuscan ambassador of Prague (the forerunner to the IAU
telegram?)  read {\it Haec immatura a me iam frustra leguntur o.y.}
[This immature female has already been read in vain by me] but
unscrambled to {\it Cynthiae figuras aemulatur mater amorum} [The
mother of love (Venus) resembles Cynthia (the Moon)]
(\cite{vanhelden89}).

The first direct detection of an Earth-sized planet around another
star would be a scientific triumph of equal significance, yet the
ultimate goal is to obtain information about its atmospheres and
surfaces, especially those properties related to its ability to
support life, or even evidence for life itself.  Spectroscopy is a
powerful technique to obtain such information
(\cite{desmarais02,seager05}), however high signal-to-noise is
required.  Regardless of whether spectroscopy is succesful, direct
detection of a planet, confirmation of its companion status, and
determination of its orbit require observations at multiple epochs.
Photometry at these epochs will produce a partial or complete phase
light curve of the planet.  Such a light curve is distinct from
photometric variability in the planet induced by its rotation and
surface features (\cite{ford01}).  The latter may be diffiult to
observe, but easy to remove by sufficiently long integration times.

\begin{figure}[h]
\begin{center}
 \includegraphics[height=3in,width=3.8in]{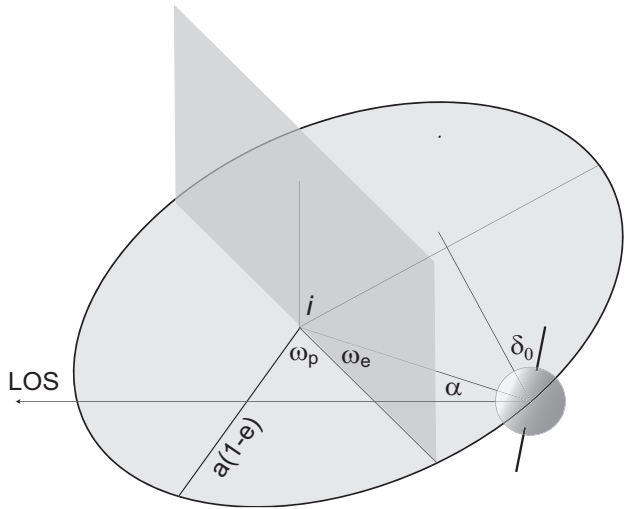}
  \caption{The geometry of the star-planet-observer system for light
  curve calculations is specified by the inclination of the normal of
  the planetary orbit plane to the line of sight ($i$), semi-major
  axis ($a$) and orbital eccentricity ($e$), the longitude of the
  periastron relative to the ascending node ($\omega_p$), the
  longitude of the spring equinox of the hemisphere most in view
  relative to the ascending node ($\omega_e$), and the obliquity
  ($\delta_0$).  $\alpha$ is the phase angle. \label{fig.geometry}}
\end{center}
\end{figure}

A theoretical model of an orbital light curve must include a full
description of the star-planet-observer geometry, including 5 orbital
parameters plus the obliquity (Figure \ref{fig.geometry}).  The light
curve will also depend on the scattering and emission properties of
the atmosphere and surface modulated by any seasonal effects induced
by finite obliquity or eccentricity.  The presence of Saturn-like
rings around a giant planet may also manifest itself by specific
photometric features (\cite{arnold04}).  The reflected and emitted
light from extrasolar giant planets has been modeled, including
effects induced by changes in effective temperature on highly
eccentric orbits
(\cite{burrows05,fortney05,dyudina05,sudarsky05,barman05}).  Although
the complete light curve of a giant planet has not been measured,
there are useful upper limits for the reflected light of close-in and
transiting planets
(\cite{charbonneau99,leigh03,green03,deming05b,snellen05}) and
detections of secondary eclipses at thermal infrared wavelengths
(\cite{charbonneau05,deming05b}).

\section{Reflected Light from Terrestrial  Planets}

\begin{figure}[h]
\begin{center}
 \includegraphics[height=2.3in,width=5in]{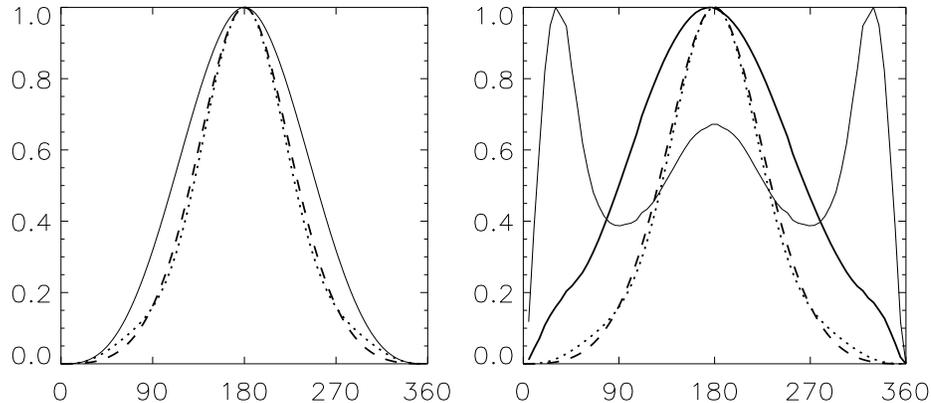}
  \caption{(a) Orbital reflected light curves of three uniform planets
  observed at $i = 90$~deg. with and different scattering laws;
  Lambertian (solid line), single RS (dotted line) and a model of
  Mercury (dashed line) (\cite{warell04}). (b) Light curves of
  ``smooth'' (ocean) planets (solid lines) with 0\% (thin) and 38\%
  (thick) cloud cover compared to RS and Mercury models. All curves
  have been normalized.
  \label{fig.scattering}}
\end{center}
\end{figure}

The reflected light from an Earth-like planet will depend on the
scattering properties (phase function) of the principle reflectors,
i.e., clouds, the atmosphere, ice, and land.  Superposed on these
effects will be seasonal variation in reflectivity, which for the
Earth is as much as 10\% (\cite{woolf02}).  However, it seems likely
that many Earth-sized planets will be non Earth-like, and it is
therefore useful to ask whether orbital light curves can distinguish
between hypothetical, broad classes of objects characterized by a
uniform reflecting surface and a simple scattering law.  The simplest
case is isotropic (Lambertian) scattering, however Solar System
surfaces are better represented by Hapke single-scattering models
(\cite{hapke93}).The phase curves of Mercury and Mars have been
successfully reproduced using a representation of the double
Henyey-Greenstein function
\begin{equation}
\label{eqn.2HG}
p(\alpha) = \frac{(1+c)(1-b^2)}{2\left(1-2b \cos \alpha + b^2\right)^{3/2}} + \frac{(1-c)(1-b^2)}{2\left(1+2b \cos \alpha + b^2\right)^{3/2}},
\end{equation}
and parameter values $b = 0.21, c = 0.7$, and $b = 0.18, c=1.1$,
respectively (\cite{warell04}).  The reflected light from a planet
with a deep, transparant atmosphere and a dark surface, e.g. a global
ocean, will be dominatd by Rayleight scattering (RS).  Single RS has
the phase function
\begin{equation}
\label{eqn.rs}
p(\alpha) = \frac{3}{4}\left(1+ \cos^2 \alpha \right).
\end{equation}
Can Hapke- and Rayleigh-scattering planets be distinguished solely on
the shape of their reflected light curves (absolute albedos require
observations in the infrared)?  Figure \ref{fig.scattering}a shows the
light curves of hypothetical Lambertian-, single Rayleigh-, and
Hapke-scattering planets with uniform surfaces as seen at an orbital
inclination angle of 60~deg.  The opposition effect is clearly
discernable, but the small difference between the Hapke- and
Rayleigh-scattering models are unlikely to be distinguishable at
observable phase angles.  The signal from planets with smooth
surfaces, e.g., a cloud-free ocean world \cite{leger04}) with a thin
atmosphere, or an ice-covered planet, will include significant
specular reflection as well as diffuse scattering.  The light curves
of such objects can be clearly distinguished (Figure
\ref{fig.scattering}b) from those of the previous cases, although
increasing cloud cover and Rayleigh scattering by an atmosphere will
moderate these effects (Williams et al., in prep).

\section{Emitted Light from Terrestrial Planets \label{section.emitted}}

The simplest terrestrial planet case is that of a rotating rocky
planet lacking oceans or atmosphere, having a uniform albedo and
blackbody emissivity, e.g., Mercury.  The thermal inertia $j$ of a
solid surface subject to diurnal heating with frequency $\Omega$ is
$\sqrt{k\rho c}$, where $k$ is the thermal conductivity, $\rho$ the
density, and $c$ is the specific heat capacity of the material.  Solid
basalt ($k = 2$~W~m$^{-1}$K$^{-1}$, $\rho = 2900$~kg~m$^{-3}$, $c =
1$~kJ~kg$^{-1}$~K$^{-1}$) has $J \approx 2400$~J
m$^{-2}$K$^{-1}$~sec$^{-1/2}$.  Dust-covered surfaces on Mars have $J$
values as low as 30~W m$^{-2}$K$^{-1}$~sec$^{-1/2}$ (\cite{mellon00}).
The thermal inertia of the lunar regolith is $\sim$45~W
m$^{-2}$K$^{-1}$~sec$^{-1/2}$, and the large main-belt asteroids have
still lower $J$ values (\cite{muller98}).  The thermal response of a
planet can be divided into two regimes, $t_{day} \ll t_{thermal}$ and
$t_{day} \gg t_{thermal}$, where the thermal response time is
\begin{equation}
\label{eqn.time}
t_{thermal} = \frac{\pi J^2}{4\sigma^2\bar{T}^6},
\end{equation}
and $\sigma$ and $\bar{T}$ are the Stefan-Boltzmann constant and the
mean temperature, respectively.  For a bare rock surface at 255~K (the
emission temperature of the Earth), $t_d$ is 2 months (note the
extreme sensitivity to $\bar{T}$).  A planet with the thermal
properties of the lunar surface has $t_{thermal} = 30$~minutes!

\begin{figure}[h]
\begin{center}
 \includegraphics[height=2.3in,width=5in]{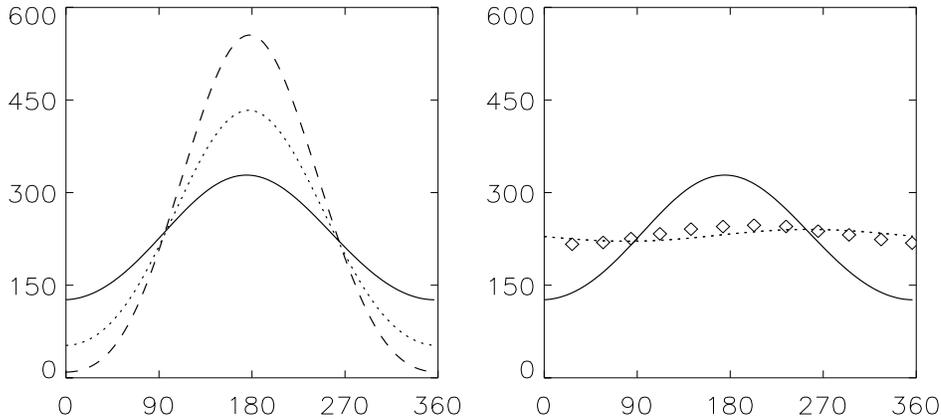}
\caption{(a) Disk-averaged emitted light curves of planets W~m$^{-2}$)
seen at $i = 60$~deg.  and with negligible thermal inertia over the
orbital period, large thermal inertia over the rotation period, and
$\delta_0 =$ 23.45 (solid line), 40, and 60 deg.  (b) Earth's emitted
light curve modeled by a 2-D energy balance model (dotted line) and a
3-D global circulation model (points) (\cite{gaidos04}) compared to
that of a simlar body with negligible thermal inertia (solid
line).\label{fig.emitted}}
\end{center}
\end{figure}

A planets with a fine-grained regoliths and no atmospheres will
experience large diurnal temperature variation, and, depending on $i$,
will exhibit significant orbital variation in disk-averaged flux.  In
contrast, a planets covered with bare rock, e.g. one on which geologic
resurfacing continues, will have a relatively small day-night surface
temperature difference and orbital variation in its emitted flux would
occur only if it had a significant obliquity and seasonality.  The
modest average obliquity of the Earth ($\delta_0 = 23.45$~deg. may not
be typical of terrestrial planets (\cite{laskar93}): The obliquities
of Mars and Venus are thought to have undergone large excursions over
Solar System history and numerical simulations of the final stage of
terrestrial planet formation produce planets with an isotropic
distribution of primordial obliquities due to the final stochastic
accretion of a few large embryos (\cite{agnor99}).  Large $\delta_0$
leads to large seasonal and hemispherical differences in incident
stellar radiation and a high amplitude light curve (\cite{gaidos04})
(Figure \ref{fig.emitted}a).

An ocean and/or an appreciable atmosphere will profoundly affect the
emitted flux from a terrestrial planet.  The relation between outgoing
infrared flux and temperature $I(T)$ for a greenhouse atmosphere
deviates significantly from that of a blackbody.  Oceans and
atmospheres transport heat along thermal gradients and have large
thermal inerties.  The energy-balance equation governing the surface
temperature $T$ of a planet as a function of time $t$ and latitude
$\theta$ can be modeled using the heat transport equation
\begin{equation}
\label{eqn.energybalance}
C\frac{\partial T}{\partial t} = S(1-A) - I(T) + \frac{1}{\cos
\theta}\frac{\partial}{\partial \theta} \left(D \cos \theta
\frac{\partial T}{\partial \theta} \right),
\end{equation}
where $S$ is the incident stellar flux, $A$ is the top-of-the
atmosphere albedo, and $C$ and $D$ parameterize the effective heat
capacity of the atmosphere and surface, and meridional heat transport,
respectively.  The net effect is to obviate diurnal variation and
attenuate seasonal variations (\cite{gaidos04}) (Figure
\ref{fig.emitted}b).  Even planets with extremely high obliquities
experience habitable temperature ranges over much of their surfaces as
a consequence of the modering effect of an ocean and atmosphere
(\cite{williams97,williams03}).  Thus, observation of marked infrared
variability alone suggests properties of a planet relating to
habitability; lack of an atmosphere or oceans and either a
geologically old surface or a high obliquity.  On the other hand, {\it
absence} of variability cannot be uniquely interpreted as suggesting
habitable conditions.  The planet might resemble Earth; it might also
be a barren planet with a very low obliquity, or have a thick runaway
greenhouse atmosphere like Venus.

\section{Terrestrial Planet Satellites}

The Earth's Moon is thought to have formed by a low-velocity glancing
impact with a Mars-sized body early in its history (\cite{canup04}).
The Moon (or the angular momentum of the Earth-Moon system) stabilizes
the Earth's obliquity against chaotic excursions and may be important
for habitability (\cite{laskar93}).  Numerical simulations of planet
formation show that potentially satellite-forming collisions are
common (\cite{ida97}) and it is dynamically plausible that Mercury and
Venus also had satellites but then lost them
(\cite{burns73,ward73,yokoyama99}).  Extrasolar satellites will be
unresolved from their parent planet (the Earth-Moon distance subtends
0.25~mas at 10~pc) and only the total signal of the system will be
observable.

\begin{figure}[h]
\begin{center}
 \includegraphics[height=2.3in,width=3.2in]{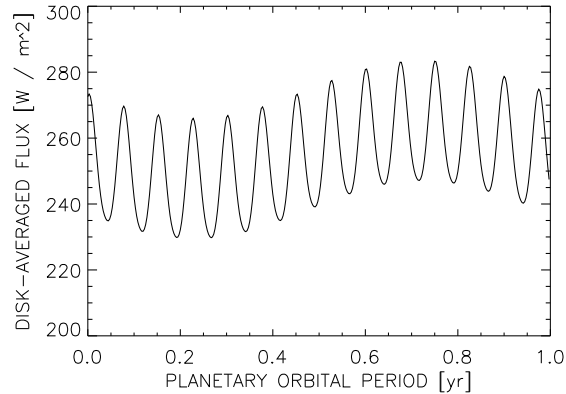}
  \caption{Disk-averaged emitted light curve of the Earth-Moon system
  (W~m$^{-2}$) seen at $i = 60$~deg.  The variation induced by the
  Moon's phases is larger than that of the
  Earth.\label{fig.satellite}}
\end{center}
\end{figure}

The theory of impact origin explains the Moon's lack of volatiles by
invoking high temperatures in the post-impact circumterrestrial disk;
this may be a common property of satellites formed in this manner.
There are two consequences of a lack of atmospheres or oceans: First,
the satellite will be darker and essentially undectable in reflected
light (the Moon-Earth flux ratio is 1.7\%).  Second, the lower thermal
inertia of the satellite (see \S \ref{section.emitted}) means that the
surface of the satellite experiences larger diurnal temperature
variation, and consequently exhibit larger flux variation at thermal
infrared wavelengths.  Despite the satellite being smaller than the
parent planet, the satellite's variation in IR flux may be larger
(Figure \ref{fig.satellite}).  Satellites of planets may be detectable
by sufficiently time-resolved measurements in the infrared.  On the
other hand, insufficiently time-resolved measurements may suffer from
confusion of the two signals and aliasing.  Satellites represent
another opportunity and a new challenge for the direct detection and
characterize terrestrial exoplanets.

\begin{acknowledgments}
Support by the NASA Terrestrial Planet Foundation Science program is acknowledged.
\end{acknowledgments}

\end{document}